\documentclass[12pt,psfig]{article}
\usepackage{epsf,graphics}
\usepackage[figuresright]{rotating}
\newcommand{\beq}{\begin{equation}}
\newcommand{\eeq}{\end{equation}}
\newcommand{\beqa}{\begin{eqnarray}}
\newcommand{\eeqa}{\end{eqnarray}}
\def\nue{{\nu_e}}
\def\anue{{\bar{\nu_e}}}
\def\numu{{\nu_{\mu}}}
\def\anumu{{\bar{\nu_{\mu}}}}
\def\nutau{{\nu_{\tau}}}
\def\anutau{{\bar{\nu_{\tau}}}}
\newcommand{\dm}{\mbox{$\Delta{m}^{2}$~}}
\newcommand{\st}{\mbox{$\sin^{2}2\theta$~}}

\newcommand{\cnv}{\mbox{$\breve{\rm C}$erenkov~}}
\def\br{{$^{8}{B} ~$}}
\def\etal{{\it et al.}}

\begin{document}
\begin{center}
{\large \bf {Progress in neutrino oscillation searches and  their
implications }}\vskip 20pt Srubabati Goswami \footnote{e-mail:
sruba@mri.ernet.in}\\ Harish-Chandra Research Institute, Chhatnag
Road, Jhusi, Allahabad - 211-019.
\end{center}

\begin{center}
Abstract
\end{center}
Neutrino Oscillation, in which a given flavour of neutrino
transforms into another is a powerful tool for probing small
neutrino masses. The intrinsic neutrino properties involved are
neutrino mass squared difference $\Delta m^2$ and the mixing angle
in vacuum $\theta$. In this talk I will summarize the progress
that we have achieved in our search for  neutrino oscillation with
special emphasis on the recent results from the Sudbury Neutrino
Observatory (SNO) on the measurement of solar neutrino fluxes. I
will outline the current bounds on the neutrino masses and mixing
parameters and discuss the major physics goals of future neutrino
experiments in the context of the present picture.

\section{Introduction}
Since it was proposed in 1929 by Pauli the question whether
neutrinos are massive or not has been an intriguing issue. The
Standard Model (SM) contains only massless neutrinos in left
handed doublets. But there is no fundamental principle like the
gauge invariance which makes the photon massless --that would
forbid a mass term for the neutrinos. Most extensions of the SM
predict small but non-zero neutrino masses. A non-zero neutrino
mass would not only constitute a signal of physics beyond standard
model it would also help in probing the underlying gauge symmetry
governing the interactions. Moreover because of the important role
played by neutrinos in stellar evolution, supernova dynamics,
nucleosynthesis and structure formation in the early universe the
implications of massive neutrinos can be very significant for
astrophysics and cosmology.

Direct bounds on neutrino masses are obtained from purely
kinematical measurements\\ $\bullet$ $m_{\nu_e} <$ 2.2 eV
{($^{3}{H} \rightarrow ^{3}{He} + e^{-} + \bar{\nu_e}$)}
\cite{nue} \\ {$\bullet$} $m_{\nu_\mu} < $ 170 KeV {(Pion decay)}
\cite{numu} \\ {$\bullet$} $m_{\nu_\tau} < $ 15.5 MeV {(Tau
decay)} \cite{nutau}.\\ A bound  on the effective mass of the
electron neutrino can be obtained from the absence of the
neutrinoless double beta decay  process  which gives
$\langle{m_{ee}}\rangle\equiv |\sum U_{ei}^2 m_{\nu_i}|\leq 0.38
p$ eV at 95\%~ CL, where $p\sim 0.6-2.8$ denotes the uncertainty
in nuclear matrix element \cite{dbd}. $U$ is the neutrino mixing
matrix and $m_{\nu_i} (i=1,2,3)$ denote the neutrino mass values.
This process  is lepton number violating and the occurrence of
such an event would indicate that neutrinos are Majorana fermions
{\it i.e.} their own antiparticles. An important bound on neutrino
mass comes from cosmology\\ \beq \sum_i m_{\nu_i} < 100 h^2 {\rm
eV} \eeq which implies masses much smaller than the kinematical
bounds.

Very small neutrino mass squared differences can be probed by
neutrino oscillation which arises if neutrinos have mass and
different flavours mix among themselves. Thanks to painstaking
experimental efforts we now have strong evidences in favour of
neutrino oscillation coming from the measurement of solar and
atmospheric neutrino fluxes. The high statistics SuperKamiokande(SK)
and SNO experiments have established the conjecture of neutrino
flavour conversion as a reality and provided important information
on neutrino mass and mixing. The purpose of this talk is to
discuss the recent developments in the neutrino oscillation
experiments and the bounds on neutrino mass squared  differences
and mixing angles obtained from these. Since the results and
implications of the SK experiment has been discussed in detail by
Mark Vagins in these proceedings the emphasis on this paper will be on the
recent SNO results. In section 2 we develop briefly the formalism
for neutrino oscillation in vacuum as well as in matter. In
section 3 we discuss the neutrino oscillation experiments and
discuss the constraint that one obtains on \dm and $\theta$ from
two and three generation oscillation analyses. In section 4 we
discuss if the current data requires a fourth sterile neutrino or
not. In section 5 we discuss the goals and salient  features
of the future experiments
which are soon expected to
give data/start operation. Finally we present the concluding
remarks.

\section{Neutrino Oscillation In Vacuum}
We consider a neutrino flavor state $|{\nu_{\alpha}}>$ created in
weak interaction processes. In general this is a superposition of
neutrino mass eigenstates $|\nu_{i}>$ \beq |\nu_{\alpha}> =
\Sigma_{i} U_{\alpha i} |\nu_{i}> \label{e1} \eeq where $U$ is the
unitary mixing matrix analogous to the Cabibbo-Kobayashi-Maskawa
matrix in the quark sector.
Assuming ultra-relativistic neutrinos with a common definite
momentum $p$, $E_{i} \simeq p + {m_{i}^2} /{2p}$i, $U$ to be real
{\it i.e.} neglecting CP violating phases in the lepton sector
the probability that a neutrino of flavour $\alpha$ gets converted
to a neutrino of flavour $\beta$ after traversing a distance $L$
is \beq P(\nu_{\alpha},0 ; \nu_{\beta},t) = \delta_{\alpha \beta}
- 4~ \Sigma_{j > i}~ U_{\alpha i} U_{\beta i} U_{\alpha j}
U_{\beta j} \sin^{2}\left(\frac{\pi L}{\lambda_{ij}}\right)
\label{npr} \eeq where $\lambda_{ij}$ is defined to be the
neutrino vacuum oscillation wavelength given by, \beq \lambda_{ij}
= (2.47 {\rm m}) \left(\frac{E}{MeV}\right)
\left(\frac{\Delta_{ij}}{eV^2}\right)^{-1} \label{wv} \eeq which
denotes the scale over which neutrino oscillation effects can be
significant; \\ $\Delta_{ij} = \mid{m_{j}^2 - m_{i}^2}\mid$. The
oscillatory character is embedded in the $\sin^{2}(\frac{\pi
L}{\lambda_{ij}})$ term. If $\Delta_{ij}$ is such that the
corresponding  $\lambda_{ij}
>> L$, the oscillatory term $\sin^2{\pi L/\lambda_{ij}}
\rightarrow 0 $, whereas $\lambda_{ij} << L$ would imply a large
number of oscillations and consequently the $\sin^2{\pi
L/\lambda_{ij}}$ term averages out to 1 /2, when integrated over
energy and/or source~/detector position.  \\ For two-generations,
$U$ can be parametrised by a single mixing angle $\theta$ as: \beq
U = \pmatrix{\cos{\theta} & \sin{\theta} \cr -\sin{\theta} &
\cos{\theta} \cr} \eeq and the conversion probability is
\begin{equation}
P_{\nu_{\alpha}\nu_{\beta}} = \sin{^2}2\theta
\sin{^{2}}(1.27{\Delta m}^{2}L/E) \label{p2nu}
\end{equation}
where \dm is in eV$^2$, $L$ is in meters and $E$ is in MeV. From
the above equation we see that oscillation effect is maximum for
$\dm \sim L/E$. Consequently the values of \dm that can be
explored in an experiment depend on the energy of the neutrino
beam and the source-detector distance. Table 1 shows the typical
$L/E$ and the \dm that can be probed in various experiments using
different neutrino sources. Since the energy usually has a spread,
each experiment is actually sensitive to a range of \dm.

\section{Matter-Enhanced Resonant Flavor Conversion}
Wolfenstein and later Mikheyev and Smirnov \cite{msw}
pointed out that
interaction of the neutrinos with matter modifies their dispersion
relation and consequently they develop an effective mass
dependent
on the matter density. For unpolarized electrons at rest, the
forward charged current scattering of neutrinos off electron gives
rise to an effective potential \beq V_{cc} = \sqrt{2}G_Fn_e \eeq
while the neutral current interaction gives a contribution
(assuming charge neutrality) \beq V_{nc}=-\sqrt{2}G_Fn_n/2 \eeq

\begin{table}
\begin{center}
\begin{tabular}{|c|c|c|c|}
\hline Experiment & Energy(E) & Distance(L) & $\Delta m^2$  \\
\hline Reactor & 1 MeV &  100m &  $10^{-2}$ eV$^2$   \\ \hline Low
Energy & 10 MeV & 100m & $10^{-1}$ eV$^2$ \\ \cline{2-4}
Accelerator & & &  \\ \hline High Energy & 1 GeV & 1 km & 1 eV$^2$
\\ \cline{2-4} Accelerator & & &  \\ \hline Atmospheric & 1 GeV &
10,000 km & $10^{-4}$ eV$^2$ \\ \cline{2-4} Neutrinos & & &  \\
\hline Solar Neutrinos & 10 MeV & $10^{11}$ m & $10^{-10}$ eV$^2$
\\ \hline Long-Baseline & 1 MeV & 1 km & $10^{-3}$ eV $^2$ \\
\cline{2-4} Reactor & & & \\ \hline Long-Baseline & 1 GeV & 1000km
& $10^{-3}$ eV$^2$ \\  \cline{2-4} Accelerator   &    &  &   \\
\hline
\end{tabular}
\end{center}
\vskip -0.5cm
\caption[Sensitivity to $\Delta m^2$]
{\label{delmsq}Sensitivity to $\Delta m^2$  in various
oscillation experiments for vacuum  mixing}
\end{table}

Considering for simplicity two neutrino flavours $\nu_e$ and
$\nu_\mu$ the neutrino propagation equation in flavour basis in
the presence of matter is \beq {i\frac{d}{dt}{\pmatrix {\nu_e \cr
\nu_\mu \cr}} ={M_F}{\pmatrix {\nu_e \cr \nu_\mu \cr}}} \eeq \beq
M_F ={U} \pmatrix {m_1^2/2E & 0 \cr 0 & m_2^2/2E \cr }
     U^{\dagger}  + EI \\ \\ \\
+ \pmatrix {{\sqrt{2} G_F n_e} & 0 \cr 0 & 0 \cr} - {\frac{G_F
n_n}{\sqrt{2}}} I \eeq The last two terms denote the matter
contribution. The terms proportional to the identity matrix
contributes to the overall phase and dropping these terms the mass
matrix in flavour basis takes the form \beq M_F =
\frac{1}{2E}{\pmatrix {\Delta m^2 sin^2\theta +{2\sqrt{2}G_F n_e
E} & \Delta m^2 sin2\theta \cr \Delta m^2 sin2\theta & \Delta m^2
cos^2\theta \cr}} \eeq $\Delta m^2 = m_2^2 - m_1^2$. The mass
eigenvalues in matter is obtained by diagonalizing the above
matrix. The  flavour and mass states in matter are now related as
\beq \pmatrix{\nue \cr \nu_\mu \cr} = \pmatrix {cos\theta_M &
sin\theta_M \cr -sin\theta_M & cos\theta_M \cr}
\pmatrix{\tilde{\nu_1} \cr \tilde{\nu_2} \cr} \eeq where the
mixing angle in matter is given as
\begin{equation}
\tan 2\theta_{M} =  \frac{\Delta m^2 \sin 2\theta}{\Delta m^2 \cos
2\theta - 2\sqrt{2}G_{F}n_{e}E} \label{thetam} \eeq $n_e$ is the
electron density of the medium and $E$ is the neutrino energy. Eq.
(\ref{thetam}) demonstrates the resonant behavior of $\theta_M$.
Assuming $\Delta m^2  > 0$ the mixing angle in matter is maximal
(irrespective of the value of mixing angle in vacuum) for an
electron density satisfying,
\begin{equation}
2\sqrt{2}G_{F}n_{e,res}E = \Delta m^2 \cos 2\theta \label{neres}
\end{equation}
This is
the Mikheyev-Smirnov-Wolfenstein (MSW) effect of matter-enhanced
resonant flavor conversion.

\section{Oscillation Experiments}
Experimental searches for neutrino oscillation can be classified into
two categories \\
$\bullet$ {Disappearance Experiment -- in which one looks for the diminution
of the neutrino flux due to oscillation to some other flavor to
which the detector is not sensitive.} \\
$\bullet$ {Appearance Experiment -- in which one looks for a new neutrino
flavor  not present in the initial beam,
which can arise from oscillation.}

\subsection{Reactor And Accelerator Neutrino Experiments}
Nuclear reactors provide an intense source of ~$\overline{\nu_e}$
with energies $\sim$ 1 MeV. Because of this low energy spectrum,
reactor based oscillation experiments are suitable for searching
for $\overline{\nu_e}$ going to any flavor using disappearance
technique.  The dominant systematic uncertainties come from the
strength of the neutrino source, the detector efficiency, the
cross-section for neutrino interactions etc. These limit the
sensitivity of the mixing angle that can be probed in these
experiments.  The systematic uncertainties are largely eliminated
when energy spectra measured at two different detector positions
are compared.

There are two types of accelerator neutrino beams. Low-energy
accelerators (the Meson-factories) provide an equal admixture of
$\nu_\mu$, $\nu_{\bar{\mu}}$ and  $\nu_e$ from decays of stopped
$\pi^{+}$ and $\mu^{+}$. These are of energy $\sim$ 10 MeV.
High-energy accelerators produce $\nu_\mu$ ($\anumu$) beams coming
from K or $\pi$ decay, the $\nu_e$ component being small. The mean
energy of these type of neutrinos are $\sim$ 1 GeV.
 These are suitable to look for $\numu$ ($\anumu$) $\rightarrow \nue$
($\anue$) or $\numu$ ($\anumu$) $\rightarrow \nutau$ ($\anutau$)
by appearance method or to measure the probability of $\numu$
($\anumu$) $\rightarrow$ $\numu$ ($\anumu$) by disappearance
technique. For the appearance experiments one needs to understand
the backgrounds thoroughly in order to identify an event arising
from oscillation. Table 2 shows the characteristics of some of the
important reactor and accelerator based experiments.

\begin{table}[t]
   \begin{center}
     \begin{tabular}{||c|c||c|c|l|l||} \hline \hline
      &  & probes & signal & & \\
      & experiment & oscillation & for & \dm in eV$^2$ & \st \\
      &           &  channel & oscillation & & \\ \hline\hline
      & G\"{O}SGEN &
          $\nue \rightarrow \nu_x$ & negative & $<0.02$ & $<0.02$ \\
      & Krasnoyarsk
        & $\nue \rightarrow \nu_x$ & negative & $<0.014$ & $<0.14$ \\
      $I$
      & BUGEY
        & $\nue \rightarrow \nu_x$ & negative & $<0.01$ & $<0.2$ \\
      & CHOOZ
        & $\nue \rightarrow \nu_x$ & negative & $<0.002$ &$<0.1$  \\
      & Palo Verde
        & $\nue \rightarrow \nu_x$ & negative & $<0.002$ & $<0.6$
      \\ \hline \hline
      & CDHSW
        &$\numu \rightarrow \nu_x$ & negative & $<0.23$ & $<0.02$ \\
      & CHARM
        &$\numu \rightarrow \nu_x$ & negative & $<0.29$ & $<0.2$ \\
      & CCFR
        &$\anumu \rightarrow \bar\nu_x$ & negative & $<15$ & $<0.02$ \\
      & E776
        &$\anumu \rightarrow \bar\nu_x$ & negative & $<0.075$
        & $<0.003$ \\
    & E531
        &$\numu \rightarrow \nutau$ & negative & $<0.9$ & $<0.004$ \\
      $II$
      & E531
        &$\nue \rightarrow \nutau$ & negative & $<9$ & $<0.12$ \\
      & LSND &$\anumu \rightarrow \anue$ &
         \bf{positive} & & \\
      & LSND &$\numu \rightarrow \nue$ &
         \bf{positive} & \raisebox{1.5ex}[0pt]
         {\bigg\} 1.2}& \raisebox{1.5ex}[0pt]{\bigg\}0.003}\\
      & KARMEN2 &$\anumu \rightarrow \anue$ & negative &
         $<0.007$&$<0.0021$ \\
       \hline \hline

      \end{tabular}
     \end{center}
\vskip -0.5cm
     \caption[The reactor and accelerator experiments]
     {\label{artable}The  major reactor (I) and accelerator (II) experiments
that are running/completed. From ref. \cite{sthesis}.}
\end{table}

With the exception of the LSND experiment  all the reactor and
accelerator  based experiments of Table 2  are seen to consistent
with no neutrino oscillation and provide exclusion regions in the
\dm - \st parameter space. In Table 2
 We display the minimum value
of \dm  and minimum value of \st
that are excluded by the experiments
that observe null oscillations.
The LSND collaboration \cite{lsnd}
 gives
$P_{\overline{\nu}_{\mu}\overline{\nu}_e}$ of
$({0.31}^{+0.11}_{-0.10} \pm 0.05)$\%  corroborating their earlier
result. They have also looked for $\nu_\mu - \nu_e$ oscillations.
This gives an oscillation probability of $(0.26 \pm 0.1 \pm
0.05)$\%. However a similar experiment KARMEN at the Rutherford
laboratories searching for $\anumu - \anue$ oscillations did not
find any evidence of oscillation \cite{karmen2}. Combining the
positive result of LSND with the no-oscillation result from
KARMEN, the earlier $\anumu-\anue$ experiment E776 at BNL
\cite{e776}, and the reactor experiment Bugey results in a very
narrow range of allowed \dm from 0.4 - 2 $eV^2$ at 95\% C.L..

\subsection{Atmospheric Neutrinos}
The primary components of the cosmic-ray flux interact with the earth's
atmosphere producing pions and kaons which can decay as:
\begin{center}
$\pi^{\pm}(K^{\pm}) \rightarrow \mu^{\pm} +
\nu_{\mu}(\overline{\nu}_{\mu})$
\end{center}
\begin{center}
$\mu^{\pm} \rightarrow e^{\pm} + \nu_e(\overline{\nu}_e) +$
\end{center}
From this decay chain one expects the number of muon neutrinos to
be about twice that of electron neutrinos and as many neutrinos as
antineutrinos. Atmospheric neutrino fluxes have been computed by
several authors \cite{gaisser,honda}. Due to uncertainties arising
from the primary cosmic ray spectrum, its composition, the
hadronic interaction cross-sections, etc. different computations
of the absolute $\nue~+ \anue$~ and $\numu~$+ $\anumu$~ fluxes
agree only within 20-30\%.

Measurement of atmospheric neutrino flux has been carried out
using two different techniques:  by imaging water $\breve{\rm
C}$erenkov detectors, -- Kamiokande \cite{fukuda} and IMB
\cite{imb} -- or using iron calorimeters as is done in  Fr\'{e}jus
\cite{fr}, Nusex \cite{nus} and Soudan2 \cite{soudan}. To reduce
the uncertainty in the absolute flux values these groups presented
the  double ratios $R$
\begin{equation}
R = {\frac{(\nu_\mu + \overline{\nu}_{\mu})/
(\nu_e + \overline{\nu}_e)_{\rm obsvd}}
{(\nu_\mu + \overline{\nu}_{\mu})/(\nu_e + \overline{\nu}_e)_{\rm MC}}}
\label{ratm}
\end{equation}
where {MC} denotes the Monte-Carlo simulated ratio. Different
calculations agree to within better than 5\% on the magnitude of
this  quantity.  The value of R was found to be less than the
expected value of unity in Kamiokande, IMB and Soudan2. This
discrepancy came to be known as the {\it atmospheric neutrino
problem} and an explanation to this was sought in terms of
neutrino oscillations. The results from the high statistics
SuperKamiokande experiment not only confirmed this but also
provided an independent and strong evidence in favour of neutrino
oscillation and hence neutrino mass from the measurement of the
zenith angle dependence of the data. The SK data indicate a
deficit of the muon-neutrinos passing upward through the earth
($\theta_{zenith}
> 90^{o}$). For the downward muon  neutrinos ($\theta_{zenith} <
90^{o}$) no such deficit was found.  For electron neutrino events
the ratio $N_{up}/N_{down}$ was found to consistent with
expectations. A convincing explanation of all aspects of SK data
comes in terms of $\nu_\mu -\nu_\tau$ oscillation. The 1289 day SK
data \cite{skatmlatest} give $\sin^2 2\theta > 0.88$ and
$1.6\times 10^{-3}$ eV$^2 < \Delta m^2 < 4 \times 10^{-3}$ eV$^2$
at 90\% C.L.. Pure $\nu_\mu-\nu_s$ oscillation is disfavoured at
99\% C.L.. The charged current data from SK looking for $\nu_\tau$
interactions is reported to be consistent with $\nu_\tau$
appearance at 2$\sigma$ level \cite{mark}.

\section{Solar Neutrinos}
The sun is a copious source of electron neutrinos, produced in the
thermonuclear reactions that generate solar energy.
The underlying nuclear process is:
\[
4\;p \;\; \rightarrow \; \alpha \; + \; 2\;e^+\;+\;2\;\nu_e \;+\;25
\;{\rm MeV}
\]
The above reaction is the
effective process driven by a cycle of reactions ({\it e.g.} the
$pp$-chain or the CNO cycle).
Neutrinos are produced in several stages
and those from a particular reaction have a characteristic
spectrum. The solar neutrino fluxes are calculated using
Standard Solar Models (SSM) the most popular among  these are the
ones due to
Bahcall and his collaborators \cite{ssm}.
\subsection{Solar Neutrino Experiments}
So far seven experiments have published results on measurements of
solar neutrino flux.
\\
$\bullet$ {\underline{ Radiochemical Detectors}}

The pioneering experiment is the Cl experiment at Homestake which
employs the reaction \cite{cl}
 \beq \nu_e + ^{37}{Cl} \rightarrow
^{37}{Ar} + e^{-}. \label{37cl} \eeq for detecting the neutrinos.
The threshold is 0.814 MeV and is sensitive to the $^{8}{B}$ and
$^{7}{Be}$ neutrinos.

Three experiments SAGE in Russia and GALLEX and GNO (upgraded
version of GNO) in Gran Sasso underground laboratory in Italy
employ the following reaction \cite{ga} \beq \nu _e \; + \;
^{71}Ga \; \rightarrow \;^{71}Ge\; + \;e^- \label{gaeq} \eeq This
reaction has a low threshold of 0.233 MeV and the detectors are
sensitive to the basic $pp$ neutrinos. Since the $pp$-chain is
mainly responsible for the heat and light generation in the sun,
detection of these neutrinos constitute an important step towards
establishment of the accepted ideas of solar energy synthesis.
 \\
$\bullet${\underline{ \cnv Detectors}}

The water \cnv detector KamioKande in Japan detected the \cnv
light emitted by electrons which are scattered in the forward
direction by solar neutrinos \beq \nu_x~+ ~e \rightarrow
~\nu_x~+~e. \label{nuescatt} \eeq In addition to $\nu_e$ this
reaction is also sensitive to $\nu_\mu$ and $\nu_\tau$ though with
a reduced strength.  Since Kamiokande's energy threshold (for the
recoil electron) was 7.5 MeV, it could measure only the $^{8}{B}$
neutrino flux. But one of the most important aspect was from a
reconstruction of the incoming neutrino track it for the first
time verified that the neutrinos are indeed of solar origin
\cite{kam}. The KamioKande is upgraded to SuperKamiokande which
started taking data from 1996. It has so far provided 1496 days of
data \cite{smy}.

The most recent results on solar neutrino flux measurement has
come from the heavy water \cnv detector at Sudbury Neutrino
Observatory \cite{snocc}.
There are three reactions which are used
by this experiment
\\
$\nu_e + d \rightarrow p + p +e$ ~~~~~~ (CC) \\
$\nu_x + e \rightarrow \nu_x + e $ ~~~~~~~~~~~(ES) \\
$\nu_x + d \rightarrow p + n +\nu_x $~~~~~~(NC) \\
The CC and ES reactions have a threshold of 5 MeV for the recoil
electron while for
NC the neutrino energy threshold is
s 2.2 MeV. Thus these reactions are sensitive only
to the $^{8}{B}$ neutrinos. The NC reaction is sensitive to all
neutrino flavours with equal strength and thus provides a direct
model independent evidence of the total solar $^{8}{B}$ flux.

In Table 3 we present the fluxes observed in these experiments
with respect to the SSM predictions.
The observed fluxes in all the experiments are less than the expectations from
SSM and this constitutes the essence of the solar neutrino problem.

\begin{table}
\begin{center}
\begin{tabular}{|c|c|c|}
\hline && \\ experiment & $R$ & composition  \\ && \\ \hline && \\
$Ga$ & 0.584 $\pm$ 0.039 & $pp(55\%),Be(25\%),B(10\%)$  \\ && \\
$Cl$ & 0.335 $\pm$ 0.029 & $B(75\%),Be(15\%)$  \\ && \\ $SK$ &
0.459 $\pm$ 0.017  & $B(100\%)$ \\ && \\ $SNO(CC)$ & 0.349 $\pm$
0.021 & $B(100\%)$ \\ && \\ $SNO(ES)$ & 0.473 $\pm$ 0.074 & $B$
(100\%) \\   \hline
\end{tabular}
\caption{
The observed solar neutrino rates relative to the $SSM$
predictions \cite{ssm} are shown along with their compositions
for different experiments.
}
\end{center}
\end{table}

However as more and more data accumulated it  became more focused.
If we combine the $^{8}{B}$ flux observed in SK with the Cl
experimental rate, it shows a strong suppression of the $^{7}{Be}$
neutrinos. The pp flux constrained by solar luminosity along with
the $^{8}{B}$ flux observed in SK leaves no room for the
$^{7}{Be}$ neutrinos in Ga. This vanishing of the $^{7}{Be}$
neutrinos renders a purely astrophysical solution to the solar
neutrino problem impossible and neutrino flavour conversion was
conjectured as a plausible solution. This was beautifully
confirmed by the SNO data. The $^{8}{B}$ flux measured by the CC
and ES reactions in SNO is
\begin{center}
$\Phi_{CC}^{SNO} = 1.75 \pm 0.07(stat) ^{+0.12}
_{-0.11}(sys) \times 10^6 cm^{-2} s^{-1}$
\end{center}
\begin{center}
$\Phi_{ES}^{SNO}
= 2.39 \pm 0.34 (stat)^{+0.16}_{-0.14}(sys) \times 10^6 cm ^{-2} s^{-1}$
\end{center}
The later is consistent with that observed by  the SuperKamiokande
(SK) detector \cite{sk1258} via the same reaction
\begin{center}
$\Phi_{ES}^{SK} = 2.32 \pm  0.03 (stat) ^{+0.08}_{-0.07} \times 10^6 cm^{-2}
s^{-1}$ \label{phiessk}
\end{center}
Since the CC reaction is sensitive only to $\nu_e$ and the ES
reaction is sensitive to both $\nu_e$ and $\nu_\mu/\nu_\tau$ a
higher ES flux would signify the presence of $\nu_\mu/\nu_\tau$.
The combination of SNO CC and SK ES data provides a 3.3$\sigma$
signal for $\nu_e$ transition to an active flavour (or against
$\nu_e$ transition to solely a sterile state). The total flux of
active  $^{8}{B}$ neutrinos determined from these measurements is
$5.44 \pm 0.99 \times 10^{6} cm^{-2} s^{-1}$ which is in close
agreement with the SSM value \cite{ssm}.

SNO has also published data on the CC spectrum and it gives a flat
spectrum consistent with the flat recoil electron spectrum
observed at SK. This is in contrast with the strong energy
dependence observed by the data on total rates. Apart from the
total fluxes, SK has also provided the fluxes measured at day and
night. The day-night flux difference observed at SK : \beqa
\frac{D-N}{(D+N)/2} = -0.034 \pm 0.022 \pm 0.013 \nonumber \eeqa
which is a $ 1.3\sigma$  effect. As we will see in the next
section all these different aspects of the solar neutrino data
contribute in shaping up the allowed areas in $ \Delta m^2
-\tan^2\theta$ plane.

\subsection{Bounds on Neutrino mass and mixing including the
SNO CC data}

In this section I present the allowed regions in
 neutrino mass and mixing parameters by performing a
global and unified $\chi^2$ analysis of the solar neutrino data
including the SNO CC rate.

\subsubsection{Two generation $\nu_e -\nu_{active}$ analysis}
The general expression for $\nu_e$ survival probability in an
unified formalism over the mass range $10^{-12} - 10^{-3}$ eV$^2$
and for the mixing angle $\theta$ in the range [0,$\pi/2$] is
\cite{petcov} \beqa P_{ee}&=&P_{\odot}P_{\oplus} + (1-P_{\odot})
(1-P_{\oplus}) \nonumber \\ && + 2\sqrt{P_{\odot}(1-P_{\odot})
P_{\oplus}(1-P_{\oplus})}\cos\xi \label{probtot} \eeqa where
$P_{\odot}$ denotes the probability of conversion of $\nu_e$ to
one of the mass eigenstates in the sun and $P_{\oplus}$ gives the
conversion probability of the mass eigenstate back to the $\nu_e$
state in the earth. All the phases involved in the Sun, vacuum and
inside Earth are included in $\xi$. This most general expression
reduces to the well known MSW (the phase $\xi$ is large and
averages out) and vacuum oscillation limit (matter effects are
absent and the phase $\xi$ is important) for appropriate values of
$\Delta m^2/E$. The procedure which we use for calculating
$P_{\oplus}$ and $P_{\odot}$ in MSW, vacuum as well as the
in-between quasi-vacuum (QVO) regions  where both $\xi$ and matter
effects are relevant is discussed in \cite{chitre}. We present in
Table \ref{totchi} the results of the $\chi^2$ analysis for
$\nue-\nu_{\rm active}$ oscillations for only rates analysis as
well as for global analysis including both rate and SK spectrum
data. We use data from the Cl, Ga, SK and SNO experiments as given
in Table 3 and the 1258 day SK recoil electron energy spectrum at
day and night \footnote{We incorporate only the SNO CC rate as the
SNO ES rate and the SNO CC spectrum still have large errors.}. We
show the best-fit values of the parameters \dm and $\tan^2\theta$,
$\chi^2_{\rm min}$ and the goodness of fit (GOF) for the small
mixing angle (SMA), large-mixing angle (LMA), LOW-QVO (low
\dm-quasi-vacuum), vacuum oscillation (VO) and Just So$^2$
solutions. The best-fit for the only rates analysis comes in the
VO region which is favored at 28.79\%. For the global rates +
spectrum analysis we get five allowed solutions LMA, VO, LOW, SMA
and Just So$^2$ in order of decreasing GOF. Best-fit comes in the
LMA region. In fig. 1 we plot the probabilities vs. energy for
this five solutions at the best-fit values obtained from only
rates and rates+spectrum analysis. The LMA and the LOW solution
can reproduce the flat recoil electron spectrum observed at SK
well and hence from the global rate and spectrum analysis LMA and
LOW are preferred. For the VO solution at the best-fit values from
only rates analysis there is a non-monotonic dependence on energy
which explains the rates data well. However the spectrum data
requires a flat energy spectrum and the best-fit shifts to $\sim
4.5 \times 10^{-10}$ eV$^2$. For the SMA solution also there is a
mismatch between the parameters that give the minimum $\chi^2$ for
the rates data and the spectrum data. The spectrum data prefers
value of $\tan^2\theta$ which is one order of magnitude lower than
that preferred by the rates data. This conflict increases with the
inclusion of the SNO CC rate as is seen from fig. 1 where we plot
the probabilities for the SMA region at best-fit values of pre-SNO
and post-SNO rate analysis. The Just So$^2$ solution cannot
account for the suppression of the \br flux as is evident from
fig. 1  but since it gives a flat probability for the \br
neutrinos the spectrum shape can be accounted for and the global
analysis gives a GOF of 8.1\%.

\begin{table}[htb]
    \begin{center}
       \begin{tabular}{||c||c|c|c|c|c||} \hline\hline
        &Nature of & $\Delta m^2$ &
         $\tan^2\theta$&$\chi^2_{min}$& Goodness\\
            &Solution & in eV$^2$&  & & of fit\\
          \hline\hline
 &SMA & $7.71\times 10^{-6}$&$1.44 \times 10^{-3}$ & 5.44 & 6.59\% \\
 &LMA & $2.59 \times 10^{-5}$ & 0.34 & 3.40 & 18.27\% \\
rates &LOW-QVO & $ 1.46 \times 10^{-7}$ & 0.67 & 8.34 & 1.55\%\\
 &VO& $7.73\times 10^{-11}$& 0.27 & 2.49 &28.79\%\\
&Just So$^2$& $5.38\times 10^{-12}$ & 1.29 & 19.26
&$6.57\times10^{-3}$\%\\ \hline
   &SMA & $5.28 \times 10^{-6}$&$3.75 \times 10^{-4}$ &
   51.14 & 9.22\%  \\
rates &LMA & $4.70 \times 10^{-5}$ & 0.38 &
   33.42 &  72.18\% \\
 +  &LOW-QVO & $ 1.76 \times 10^{-7}$ & 0.67 & 39.00 & 46.99\%
   \\
 spectrum   &VO& $4.64\times 10^{-10}$& 0.57 & 38.28 & 50.25\%\\
    &Just So$^2$&$5.37\times 10^{-12}$& 0.77&51.90&8.10\%\\
         \hline\hline
        \end{tabular}
      \end{center}
\vskip -0.5cm
      \caption[$\chi^2$ fits to the solar data for  $\nue-\nu_{\rm active}$]
      {\label{totchi}The best-fit values of the parameters,
        $\chi^2_{min}$, and the goodness of fit from the
       global analysis of
       rates and rates+spectrum data for MSW oscillations involving
       two active neutrino flavors \cite{sthesis}.}
\end{table}

In fig. 2 we show the allowed regions in the $\Delta
m^2-\tan^2\theta$ plane from the analysis of total rates and
rate+spectrum. The significant change in the allowed regions after
including the SNO results is the disappearance of the SMA region
even at 99.73\% C.L. ($3\sigma$) as a result of increased conflict
between the total rates and SK spectrum data \cite{bcgk}. We see
from fig. 2 that maximal mixing ($\tan^2\theta=1$) is not allowed
in the LMA region but is allowed in the LOW region. In fig. 3 we
show the allowed regions in $\Delta m^2 - \tan^2\theta$ plane by
\\ (i) using a Cl rate renormalised by 20\% upwards in view of the
fact that this is till an uncalibrated experiment\\ (ii) by using
a $^{8}{B}$ flux normalisation factor of 0.75 in view of the large
uncertainties associated with it due to the uncertainties in the
${^{7}{Be}}(p,\gamma) {^{8}{B}}$ cross-section. \\ Both these cases
gives an enlarged allowed region encompassing the maximal mixing
solution for both LMA and LOW \cite{dp2}.

\subsubsection{Energy independent solution}

In Table 5 we give the results assuming an energy independent
survival probability \beq P_{ee} = 1 - \frac{1}{2}\sin^22\theta
\eeq

\begin{table}[htb]
\begin{center}
\begin{tabular}{ccccc}
\hline\hline
&$X_{B}$&$\sin^22\theta$ ($\tan^2\theta$)
& $\chi^2_{\min}$ & g.o.f   \\
\hline
Chlorine& 1.0&0.93(0.57)& 46.06 &23.58\%\\
Observed&0.72 & 0.94(0.60) & 44.86 & 27.54\%\\
\hline
Chlorine&1.0 &0.87(0.47) &41.19& 41.83\%\\
Renormalised&0.70&0.88(0.48)&38.63 & 48.66\%\\
\hline \hline
\end{tabular}
\end{center}
\caption{ The best-fit value of the parameter, the $\chi^2_{\min}$
and the g.o.f from a combined analysis of rate and spectrum with
the energy independent solution \cite{dp2}. }
\end{table}
This solution gives a better g.o.f to the global data as compared
to the SMA solution.

\subsubsection {Three Generation Neutrino Oscillation Parameters
after SNO}

We consider the picture $\Delta m^2_{21} = \Delta m^2_\odot$,
$\Delta m^2_{31} = \Delta m^2_{CHOOZ}, \simeq \Delta m^2_{atm} =
\Delta m^2_{32}$ \\ and the mixing matrix \beqa U  & = &\pmatrix
{c_{13}c_{12} & s_{12}c_{13} & s_{13} \cr -s_{12}c_{23} - s_{23}
s_{13} c_{12} & c_{23} c_{12} - s_{23} s_{13} s_{12} & s_{23}
c_{13} \cr s_{23} s_{12} - s_{13} c_{23} c_{12} & -s_{23} c_{12} -
s_{13} s_{12} c_{23} & c_{23} c_{13} \cr} \eeqa where we have
neglected the CP violation phases. For $\Delta$m$_{31}^2
>>  \Delta$m$_{21}^2 \approx$ the matter potential in sun , the
$\nu_3$ state experiences almost no matter effect and  MSW
resonance can occur between $\nu_2$ and $\nu_1$ states. One can
show the $\nu_e$ survival probability for this case to be \beqa
P_{ee} = c_{13}^4 P_{ee}^{2gen} + s_{13}^4 \eeqa where
$P_{ee}^{2gen}$ is of the two generation form in the mixing angle
$\theta_{12}$. We present in fig. 4 the allowed areas in the 1-2
plane at 90\%, 95\%, 99\% and 99.73\%  confidence levels for
different sets of combination of $\Delta m^2_{31}$ and $\tan^2
\theta_{13}$, lying within their respective allowed range from
atmospheric+CHOOZ and solar+CHOOZ analysis. Since CHOOZ data
restricts $\tan^2\theta_{13}$ to very small values ($< 0.075$) there
is not much change in the allowed regions compared to the
two-generation plots. In the three flavor scenario also there is
no room for SMA MSW solution at the 3$\sigma$ level (99.73\%
C.L).The mixing matrix at the best-fit value of solar+CHOOZ
analysis is \cite{bcgk3}
\\
\begin{equation}
U \simeq {\pmatrix {2\sqrt{\frac{2}{11}}  & \sqrt{\frac{3}{11}} & 0
\cr -\sqrt{\frac{3}{22}} & \frac{2}{\sqrt{11}} &
\frac{1}{\sqrt{2}} \cr \sqrt{\frac{3}{22}} & -\frac{2}{\sqrt{11}}
& \frac{1}{\sqrt{2}} }}
\end{equation}
Thus the best-fit mixing matrix is one where the neutrino pair
with larger mass splitting is maximally mixed whereas the pair
with splitting in the solar neutrino range has large {\it but not}
maximal mixing.

\section{Do we need a sterile neutrino?} The three
neutrino oscillation phenomena mentioned above --  namely the
solar neutrino problem, the atmospheric neutrino anomaly and the
$\nu_\mu - \nu_e$ oscillation observed by the LSND group --
require three hierarchically different mass ranges,
\begin{eqnarray}
\Delta m_{sun}^2&\sim 5 \times 10^{-5}
\ eV^2\ ,& \sin^2 2\theta\sim 0.8
(LMA~~~ MSW)  \nonumber\\
\Delta m_{Atm}^2&\sim  3 \times 10^{-3} \ eV^2,&\  \sin^2 2\theta\sim 1 \nonumber \\
\Delta m_{LSND}^2&\sim  0.4-2\ eV^2&\  \sin^2 2\theta\sim 0.001 - 0.01 \nonumber
\label{three}
\end{eqnarray}
which cannot be accommodated in a three-generation picture.
It has been widely realised that a remedy of this situation might
be the introduction of a fourth neutrino. According to the LEP data
there are three light active neutrino species. So the fourth
neutrino has to be sterile.  Introducing an additional sterile
neutrino many schemes are possible. Detailed analysis shows that
complete hierarchy of four neutrinos ($m_1 << m_2 << m_3 << m_4$)
is not favoured by current data \cite{grimus}. Mass patterns with
three neutrino states closely degenerate in mass and the fourth
one separated from  these by the LSND gap (the 3+1 scheme) is also
not preferred \cite{valle}. The 2+2 scheme in which two degenerate
mass states are separated by the LSND gap  is also disfavoured as
it requires dominant oscillation to sterile neutrinos  for either
solar or atmospheric neutrinos. The still allowed scheme is the so
called mixed 2+2 scheme in which the  solar neutrino oscillation
is due to $\nu_e$ going to a mixture of $\nu_\tau$ and $\nu_s$
and atmospheric neutrino problem can be explained by $\nu_\mu$
transition to a mixed state containing both $\nu_\tau$ and $\nu_s$
\cite{gar2}.

\section{Goals for future experiments}
In the context of the present picture it is worthwhile at this
point to discuss what should be the major aims for the future
experiments. These can be divided into two categories

\subsection{Independent confirmation of the existing anomalies}
\underline{Solar Neutrinos}\\ The hints of neutrino oscillation
coming from the earlier experiments is now established on a firm
footing by the SK and SNO data. The main goal for the future
experiments is discrimination between the LMA and LOW regions
which are still allowed. Evidence in support of the LMA region is
supposed to come from the  KamLand experiment in Japan
\cite{kamland}. This  is the first terrestrial experiment to probe
the solar neutrino anomaly. It will look for oscillation by
studying the flux and energy spectra of $\bar{\nu_e}$ produced by
Japanese commercial nuclear reactors. The typical energy and
length scales are $E_{\nu} \sim $ 3 GeV, L $\sim$ 3 $\times
10^{5}$ m , $\Delta m^2 \sim 10^{-5}$ eV$^2$. Thus KamLand can
probe the L/E dependence of the oscillations in the LMA region. It
has started taking data in January 2002 and the results are
expected to come this year.

Borexino is a
300 tons of liquid scintillator detector using trimethyle
boroxine as a target \cite{borexino}. Its main focus is the
detection of 0.862 MeV solar $^{7}{Be}$ neutrinos which requires
ultra low natural radioactive levels.
A 4 tons prototype called counting test facility (CTF)
has demonstrated extremely low radioactive level
 ($10^{-16}$g/g of U/th) can be achieved.
SSM predicts $\sim$ 55/day in Borexino while SMA, LMA, LOW and VO
solutions predict respectively $\sim$ 10-12/day, LMA $\sim$
24/day, LOW $\sim$ 23/day, VAC $\sim$ 10-45/day with seasonal
variations. Although LMA and LOW has almost the same number of
events Borexino  can distinguish between these two since the LOW
region gives rise to a day-night asymmetry  in Borexino due to
earth regeneration  \\ \underline{Atmospheric Neutrinos}\\
 Plans
to increase the sensitivity of the accelerator neutrino
oscillation experiments, by  extending the baseline beyond the
limits of the laboratory by sending intense neutrino beams towards
a large and far away detector are in progress. These are the
proposed long baseline experiments. These can have  a sensitivity
down to \dm $\approx 10^{-3}$ eV$^2$ and using these one can {\it
cross-check the atmospheric neutrino results using neutrino beams
of well known properties}. K2K is the first Long baseline
experiment to declare its data \cite{k2k} and it reports a
positive evidence of oscillation as we have heard from the
previous speaker. The P-875 (MINOS) project plans to send a
$\nu_{\mu}$ beam from FNAL to the Soudan mine \cite{minos} with a
baseline of 730 km. It will have maximum sensitivity to $\nu-\mu
-\nu_\tau$ and $\nu_\mu-\nu_e$ oscillations in the parameter space
suggested by the SK atmospheric neutrino data. In Europe one
proposed project is to send a beam from CERN to the
  ICARUS detector in the Gran Sasso underground
laboratory in Italy with $L$ = 732 km \cite{cg} and explore
$\nu_{\mu} - \nu_{\tau}$ and $\nu_{\mu} - \nu_e$ oscillations.

There is a proposal of a massive magnetized tracking calorimeter
(MONOLITH) for detecting the atmospheric muon neutrinos and
directly observing the $L/E$ pattern \cite{monolith}. \\
\underline{LSND}\\ Independent confirmation of LSND results are
expected to come from the Miniboone experiment at Fermilab which
is scheduled to start collecting data  soon \cite{miniboone}.

\subsection{Precise determination of the oscillation parameters}
The SK and SNO data has provided the long awaited evidence in
favour of neutrino oscillation. The future experiments are now
geared towards determining the oscillation parameters accurately
and determine the leptonic CKM matrix. Important role  will be
played by the long baseline experiments discussed earlier and by
the proposed  neutrino factories \cite{nufac}. These will use muon
storage rings to produce intense neutrino  beams. A 20-30 GeV
neutrino factory with $10^{19}$ muon decays/year provide what is
known as an entry-level machine whereas a 50  GeV neutrino factory
providing  10$^{20}$ muon decays/year is termed an
high-performance machine. Below I list which parameters can be
determined to what precision from these experiments.
\\
Kamland experiment can tell us how close ($\pm$ few \%)
$\sin^22\theta_{12}$
 to 1 and also it is possible to determine of
$|\Delta m^2_{21}|$ precisely (to within 10\%) if LMA solution is
the correct one.  $\sin^22\theta_{13}$ can be determined with a
precision of $\sim$ 0.01 (long baseline experiments), $0.001$ (
entry level neutrino factory ), $\sim 10^{-4}$ ( high performance
neutrino factory). $\sin^22\theta_{23}$  can be determined to an
accuracy of 10\% (long baseline accelerator experiments and entry
level neutrino factories) and in high performance neutrino
factories this can be improved to $<5$\%.  $\Delta m^2_{23}$  can
be determined by long baseline accelerator experiments and entry
level neutrino factories to an accuracy of $\sim$ 10\%. and in
high performance neutrino factories it can be determined to within
$<$ 1\%. Sign of $\Delta m^2_{23}$  (normal or inverted hierarchy)
can be determined from studying the matter effects on neutrino
propagation in long baseline accelerator experiments and Neutrino
Factories. CP violation in the lepton sector can be probed in
neutrino factories, A new tritium beta decay experiment KATRIN is
being planned to have a sensitivity limit of $\sim$ 0.3 eV
\cite{katrin}. This can provide the complimentary information on
the absolute neutrino mass scale. Neutrino less double beta decay
experiments  are planned to have an increases sensitivity to
$\langle{m_{ee}}\rangle$. These are CURORE ($\sim 0.1 $ eV), MOON
($\sim 0.03$ eV), GENIUS ($\sim 0.002 $ eV) \cite{newnbd}.

\section{Concluding Remarks}
Last twenty years have seen a substantial progress in finding a
definite answer to the question of neutrino mass and neutrino
oscillation has emerged as a powerful probe to explore small
neutrino masses. We now have two conclusive evidences in favour of
neutrino flavour conversion coming from the \\ $\bullet$ The
up/down asymmetry of the atmospheric muon neutrinos observed by
SuperKamiokande  signifies dominant $\nu_\mu - \nu_\tau$
conversion.\\ $\bullet$ Combination of SuperKamiokande and the SNO
charged current (CC) data signals the presence of
$\nu_\mu$/$\nu_\tau$ in the solar $\nu_e$ flux at more than
3$\sigma$ level.
\\
$\bullet$ The most plausible and comprehensive
 explanation of both atmospheric and
solar neutrino problem comes in terms of neutrino oscillation. The
atmospheric neutrino anomaly indicates $\Delta m^2 \sim 10^{-3}$
eV$^2$ and $\tan^2\theta \sim 1$. The solar neutrino data gives a
best-fit $\Delta m^2 \sim 5 \times 10^{-5}$ eV$^2$ and
$\tan^2\theta \sim 0.38$. Thus both atmospheric and solar neutrino
problem indicate large mixing angles.
\\
$\bullet$ The combined effect of SNO CC data and the SK recoil
electron spectrum data disfavours SMA solution to the solar
neutrino problem and no allowed region is obtained in this region
at 3$\sigma$ level from the global analysis.
\\
  $\bullet$ A three
generation analysis involving solar, atmospheric and CHOOZ data
indicate a small value of $\tan^2\theta_{13}$ ($<0.075$) which is
the common mixing angle connecting both solar and atmospheric
sectors. Because of this the allowed regions in $\Delta m^2_{12}$
and $\theta_{12}$ for solar and $\Delta m^2_{23}$ and
$\theta_{23}$ for atmospheric remain almost the same as in the two
generation case. \\
 $\bullet$ If
the LSND data of positive evidence of neutrino oscillation is
taken to be true then one requires the presence of a sterile
neutrino. Of many models involving sterile neutrinos only the
mixed 2+2  models are still consistent with the current data on
solar and atmospheric neutrinos. \\ $\bullet$ Future experiments
are planned to establish the oscillation solution by observing the
actual oscillation pattern. \\ $\bullet$ With the conventional 
long baseline
and neutrino factory experiments neutrino physics will enter the
era of precision measurements. \vskip 20pt Note Added: This talk
was given in January 2002. The global analyses presented in the
talk include the SNO CC data and the 1258 day SK spectrum data.
Since then we have the 1496 day zenith angle spectrum data
\cite{smy} and the SNO NC data \cite{snonc,snodn}.  The main
results  are 
(i) The total flux measured with the NC reaction is consistent with the 
SSM value.
(ii)Comparison of the SNO NC and the CC
results establishes an active non-electron flavour component in
the solar $\nu_e$ flux at more than 5$\sigma$ level.(iii)The
probability of the LOW region decreases considerably as a combined
effect of including the SNO NC data as well as the 1496 day SK
spectrum data.

\vskip 20pt
I thank the organizers of whepp-7 for inviting me to give this talk.
I would like to acknowledge my collaborators A. Bandyopadhyay, S. Choubey,
A.Joshipura, K. Kar, D.Majumder, D.P. Roy and A. Raychaudhuri.

\begin{figure}
\vskip -1.9in
\centerline{\epsfxsize=0.9\textwidth\epsfbox{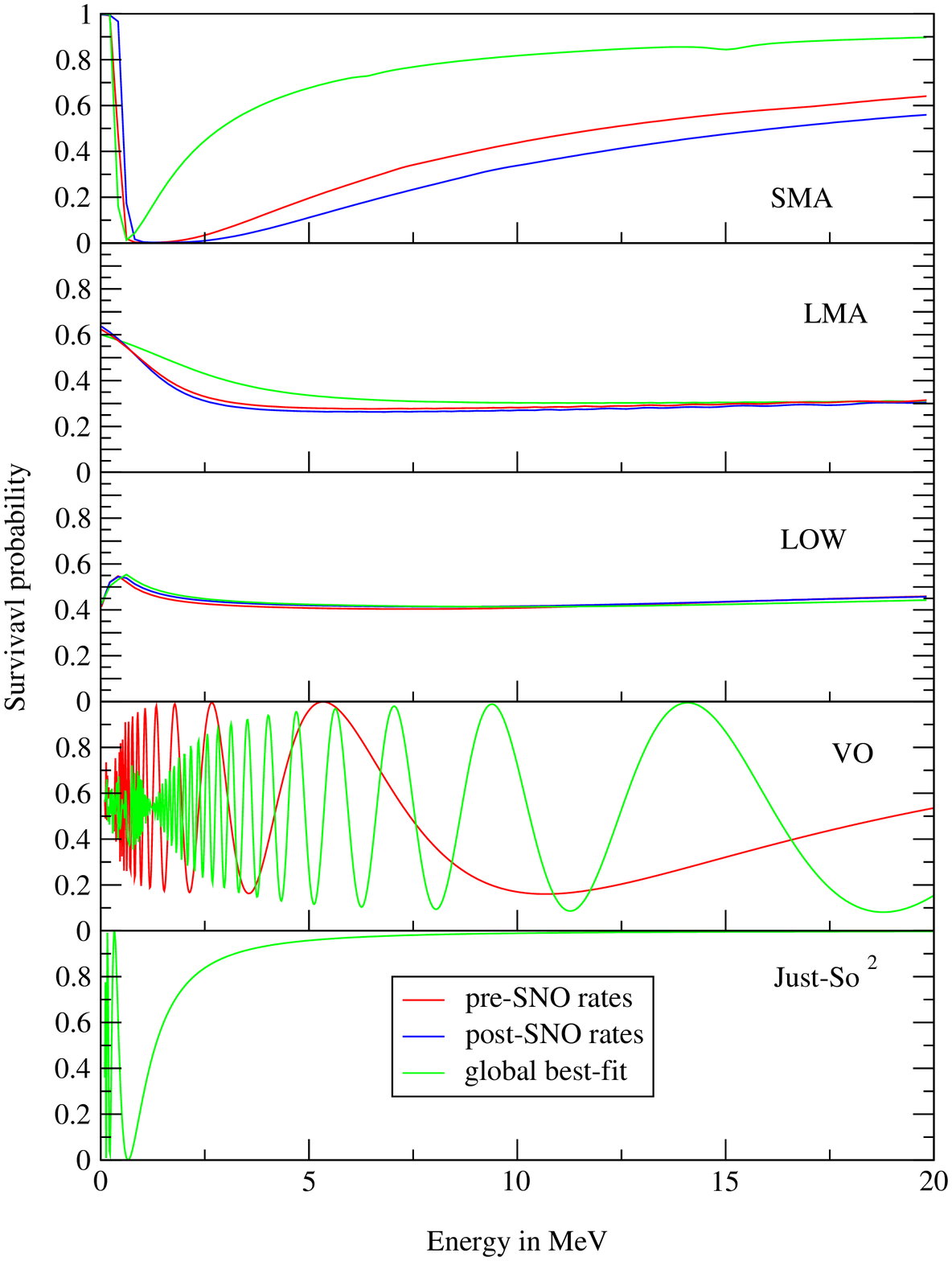}}
\caption[Probability]{\label{prob} $P_{ee}$ vs. Energy at the best-fit values 
obtained from rates and rates+spectrum analysis}
\end{figure}

\begin{figure}
\vskip -3cm
\centerline{\epsfxsize=0.9\textwidth\epsfbox{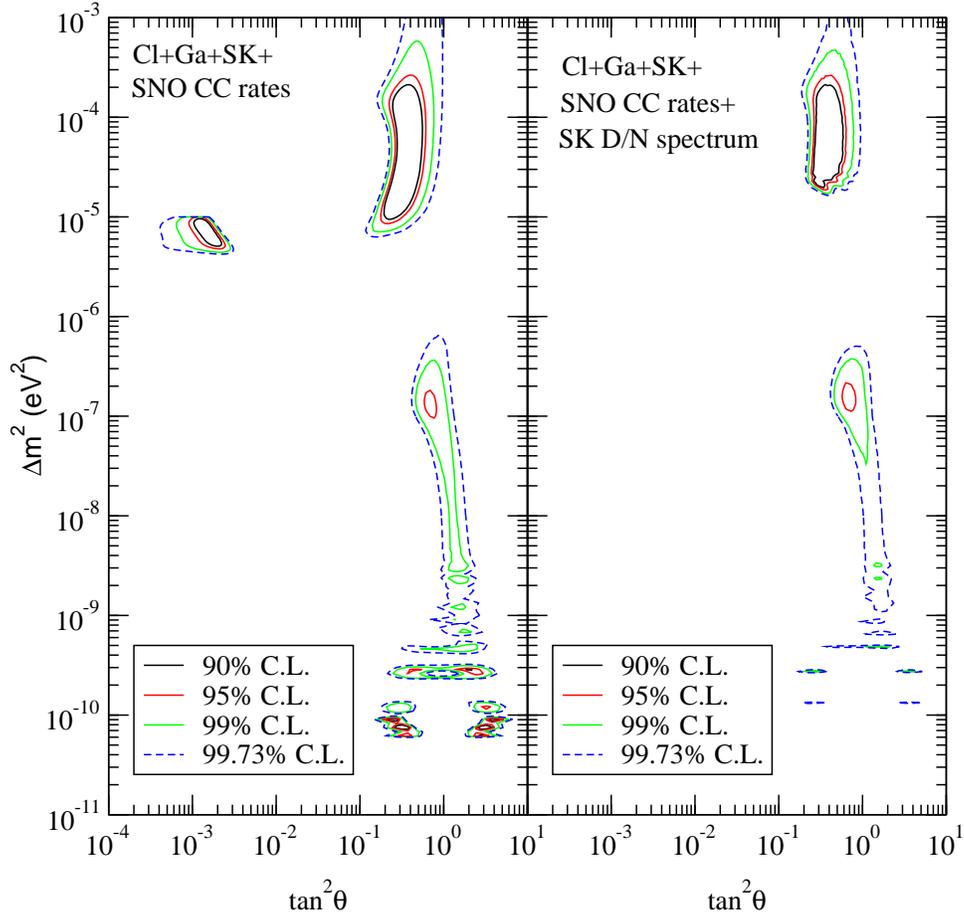}}
\vskip -1in
   \caption[The allowed areas from
   global analysis of solar data]{\label{contsolar}
   The 90\%, 95\%, 99\% and  99.73\% C.L. allowed areas from the
   analysis of the total rates (left panel) and global analysis of
the rates
   and the 1258 day SK recoil electron spectrum at day and night
(right panel),
   assuming MSW conversions to sequential neutrinos.}
\end{figure}

\begin{figure}
\topmargin -1in
\centerline{\epsfxsize=1.0\textwidth\rotatebox{270}{\epsfbox{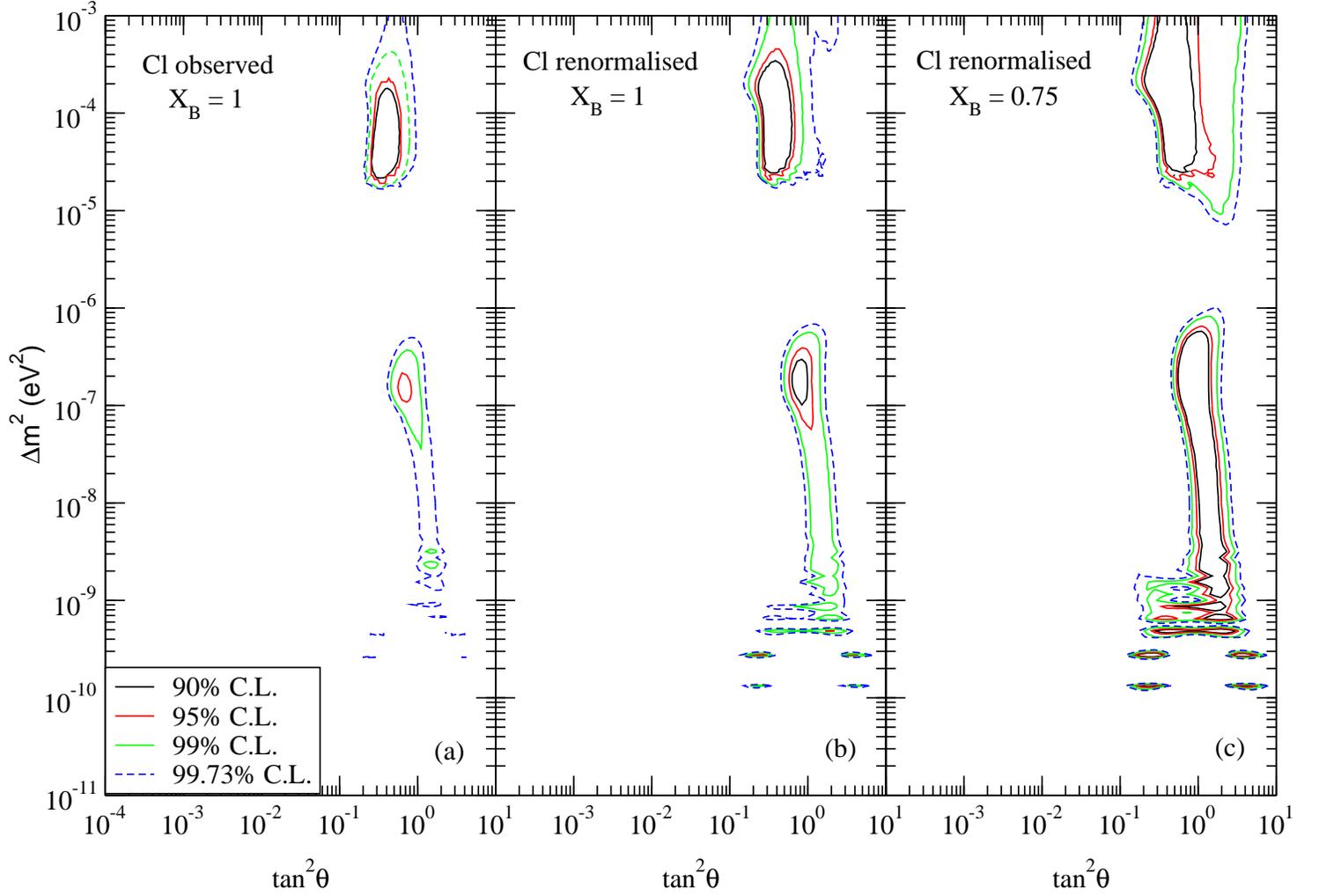}}}
\caption
{
The 90, 95 and 99 and 99.73\% C.L. allowed area from the
global analysis of the total rates from Cl (observed  and
20\% renormalised), Ga,SK and SNO (CC) detectors
and the 1258 days SK recoil electron spectrum at day and night,
assuming MSW conversions to active neutrinos.}
\end{figure}

\begin{figure}
\vskip 0.5in
\centerline{\epsfxsize=0.9\textwidth\rotatebox{270}{\epsfbox{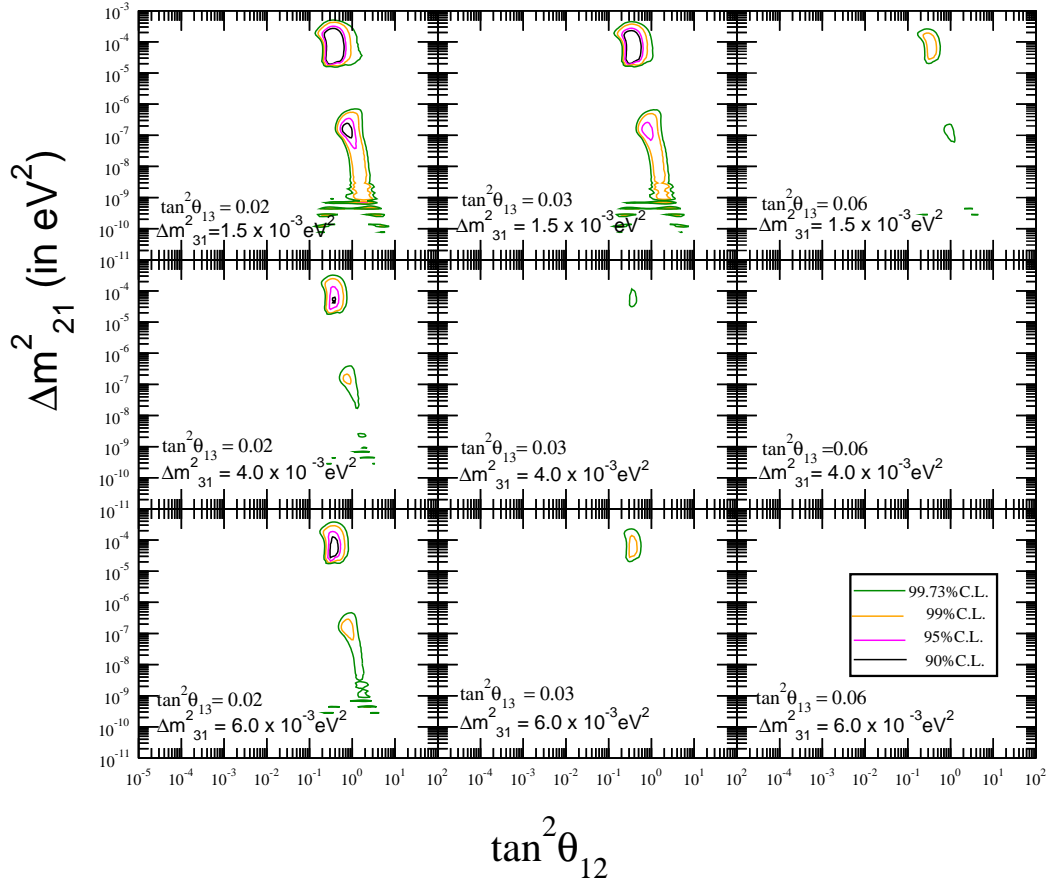}}}
\caption{The allowed areas in $\tan^2\theta_{12}-\Delta m^2_{21}$ plane
from   
solar+CHOOZ 
analysis for different fixed values of $\Delta m^2_{23}$
and $\tan^2\theta_{13}$.} 
\end{figure}

\end{document}